%
%

\input phyzzx

\overfullrule=0pt


\def\<{\langle}
\def\>{\rangle}
\def\e{\hfill\break}


\def\PRL{Phys.~Rev.~Lett.}

\def\PL{Phys.~Lett.}

\def\NP{Nucl.~Phys.}

\def\USP{Sov.~Phys.~Usp.}

\def\PR{Phys.~Rev.}

\def\PRP{Phys.~Rep.}


\REF\SEIWITONE{
\baselineskip=18pt plus 2pt
N.~Seiberg and E.~Witten \journal \NP &B426 (94) 19;
N.~Seiberg and E.~Witten \journal \NP &B431 (94) 484.}

\REF\KLELERYAN{A.~Klemm, W.~Lerche, S.~Yankielowicz and S.~Theisen
\journal \PL &B344 (95) 169.}

\REF\KLELER{A.~Klemm, W.~Lerche and S.~Theisen, CERN preprint 
CERN-TH/95-104.}

\REF\ARGFAR{P.C.~Argyres and A.E.~Faraggi \journal \PRL &74 (95)
3931.}

\REF\DANSUN{U.H.~Danielsson and B.~Sundborg \journal \PL
&B358 (95) 273, Uppsala preprint USITP-95-12; \e

M.~Douglas and S.~Shenker \journal \NP &B447 (95) 271; \e

A.~Brandhuber and K.~Landsteiner \journal \PL &B358 (95) 73; \e

A.~Hanany and Y.~Oz \journal \NP &B452 (95) 283;\e

P.~Argyres, M.~Plesser and A.~Shapere \journal \PRL &75 (95) 1699;\e

J.A.~Minahan and D.Nemeschansky, preprint USC-95-019;\e

M.~Matone \journal \PL &B357 (95) 342;\e

K.~Ito and S.-K.~Yang \journal \PL &B366 (96) 165;\e

P.C.~Argyres and A.D.~Shapere, Rutgers preprint RU-95-61;\e

A.~Hanany, preprint IASSNS-HEP-95-76. 
}

\REF\ADS{I.~Affleck, M.~Dine and N.~Seiberg \journal \NP &B241 (94) 493.} 

\REF\AMAKON{D.~Amati, K.~Konishi, Y.~Meurice, G.C.~Rossi and
G.~Veneziano \journal \PRP &162 (88) 169.}

\REF\NSVZ{V.A.~Novikkov, M.A.~Shifman, A.I.~Vainshtein and V.I.~Zakharov
\journal \NP &B260 (85) 157.}

\REF\GRISIE{M.T.~Grisaru, W.~Siegel and M.~Rocek \journal \NP &B159 (79) 
429.}

\REF\SEIZERO{N.~Seiberg \journal \PL &B206 (88) 75.}

\REF\FINPOU{D.~Finnell and P.~Pouliot \journal \NP &B453 (95) 225.}

\REF\BER{C.W.~Bernard \journal \PR &D19 (79) 3013.}

\REF\COR{S.F.~Cordes \journal \NP &B273 (86) 629.}

\REF\FUC{J.~Fuchs \journal \NP &B282 (87) 437.}

\REF\SEIONE{N.~Seiberg \journal \PL &B318 (93) 469.}

\REF\SHIVAI{M.A.~Shifman and A.I.~Vainshtein \journal \NP &B359 (91)
571.}

\REF\VAIZAK{A.~Vainshtein, V.~Zakharov, V.~Novikov and M.~Shifman
\journal \USP &25 (82) 195.}

\REF\THO{G.~'t~Hooft \journal \PR &D14 (76) 3432.}

\REF\BERWEI{C.W.~Bernard, N.H.~Christ, A.H.~Guth and E.J.~Weinberg
\journal \PR &D16 (77) 2967.}

\REF\AFF{I.~Affleck \journal \NP &B191 (81) 429.}


\pubnum{
KEK-TH-470 \cr
KEK Preprint 95-207 \cr
H \cr
}

\titlepage

\title{One-Instanton Calculations in $N=2$ 
Supersymmetric $SU(N_c)$ Yang-Mills Theory} 

\author{Katsushi Ito
\footnote{\star}{E-mail address: ito@het.ph.tsukuba.ac.jp}
\address{Institute of Physics, University of Tsukuba, Ibaraki 305, Japan}}
\andauthor{
Naoki Sasakura
\footnote{\star\star}
{E-mail address: sasakura@tuhep.phys.tohoku.ac.jp}
\address{Theory Group, KEK, Tsukuba, Ibaraki 305, Japan}
\andaddress{Department of Physics, Tohoku University, Sendai 980-77, Japan}
}

\abstract{ 
We calculate the one-instanton contribution to 
the prepotential in $N=2$ supersymmetric $SU(N_c)$ Yang-Mills theory 
{}from the microscopic viewpoint. 
We find that 
the holomorphy argument simplifies the group integrations of the
instanton configurations.
For $N_{c}=3$, the result agrees with the exact solution. 
}

\endpage


\leftskip=-40pt

\hsize=6.8in


Following the works by Seiberg and 
Witten[\SEIWITONE ],
the quantum moduli space  of the $N=2$ 
supersymmetric QCD in the Coulomb branch has been 
studied extensively [\KLELERYAN - \DANSUN].
The holomorphy and duality in the low energy effective theory 
determine the prepotential exactly, which includes the non-perturbative
instanton effects.
{}From the microscopic viewpoint, 
this exact result provides a non-trivial and quantitative test 
to the  method of instanton calculations [\ADS,\AMAKON,\NSVZ].

The perturbative non-renormalization 
theorem[\GRISIE]
shows that in the $N=2$ supersymmetric Yang-Mills theory
the first non-trivial correction to the 
prepotential beyond the  one-loop effect 
is the one-instanton 
contribution[\SEIZERO].
Finnell and Pouliot [\FINPOU] compute the four-fermi interaction in 
$N=2$ $SU(2)$ Yang-Mills theory [\SEIZERO] 
and show that the amplitude is the same as
the one predicted by the exact prepotential.
In this letter, we shall perform the one-instanton calculation
for the $N=2$ supersymmetric $SU(N_c)$ Yang-Mills theory,
and determine the one-instanton correction to the prepotential.
Our result agrees with the known exact result for $N_c=3$ case 
[\KLELER]
and satisfies the various limits  required from the exact 
solution.

One of the difficulties in the direct instanton calculation in the Higgs
and Coulomb phase is the group 
integration over the instanton configurations.
This integration is done over the embedding of the $SU(2)$ to $SU(N_c)$, where
the instanton resides.
The stability group of this embedding is 
$U(1)\times SU(N_c-2)$[\BER],
while the integrand has the additional $SU(2)$ symmetry.
Hence the group integration to be performed is over
the Grassmannian manifold $U(N_c)/(U(2)\times U(N_c-2))$.

This group integration has been studied [\COR,\FUC] in the case of 
the $N=1$ supersymmetric $SU(N_c)$
QCD with $(N_c-1)$ fundamental matters in the Higgs 
phase[\ADS,\NSVZ,\COR,\FUC].
In this case, one can choose special vacuum expectation values of 
the matter scalar fields with global $SU(N_c-1)$ symmetry,  
which make the group integration easily.
For the $N=2$ $SU(N_c)$ Yang-Mills theory in the Coulomb phase, 
however, it is impossible to make such a simplification,  
since  the vacuum expectation values of the 
adjoint scalar fields  
break all the non-abelian gauge symmetries and cannot have such a global
symmetry.

One of the new points in  the present work  is to apply  the holomorphy 
argument [\SEIONE,\AMAKON,\SHIVAI]  to the  group integration.
{}From the holomorphy, the result of the group integration
should be independent of the expectation values of the
conjugate scalar fields. 
This fact allows  us  to tune these vacuum expectation values as simple
as possible, while the vacuum expectation values of the scalar
fields are kept as arbitrary.
Thus we can calculate the four point function of the classically 
massless fermions directly.  

The $N=2$ supersymmetric Yang-Mills theory contains an $N=1$ chiral
multiplet $\phi=(A,\psi)$ with the adjoint representation 
as well as  an $N=1$ vector multiplet $W_\alpha = (v_\mu,\lambda)$.
The Lagrangian is
$$
{\cal L}=2\int d^4\theta {\rm Tr}(\phi^\dagger e^{-2gV}\phi 
e^{2gV})
+{1\over 2g^2}\left(\int d^2\theta {\rm Tr} W^\alpha W_\alpha
+ {\rm h.c.}\right),
\eqn\lag
$$
where $g$ is the gauge coupling constant.
We will examine this Lagrangian in terms of component fields
in the Wess-Zumino gauge.

The potential term $(g^2/4){\rm Tr}([A,A^\dagger]^2)$ 
has the classical flat directions
$$
\eqalign{
\< A\> &=\Omega A_0 \Omega^\dagger,\cr
{A_0}_i^{\ j}&=a_i\delta_i^j\ \ 
\left(\sum_{i=1}^{N_c} a_i=0\right), \cr
}
\eqn\expval
$$
where $\Omega$ is an element of $SU(N_c)$.
For generic $a_i$'s, 
the non-abelian symmetry is completely broken 
to $U(1)^{N_c-1}$ and the system is in the Coulomb phase.

Let us first consider the case that  the scalar vacuum 
expectation values vanish.
Then the classical euclidean equation of motion of the gauge field
has instanton solutions. In the singular gauge, the instanton
solution with unit topological charge located at the origin 
is given by 
$
v_\mu = {2\over g}{\rho^2 \bar{\eta}_{a\mu\nu}x_\nu \over
x^2(x^2+\rho^2)}GJ^aG^{\dagger}
$  [\VAIZAK],
where $\bar{\eta}_{a\mu\nu}$ is 't Hooft 
$\eta$-symbol [\THO] and $\rho$
is the instanton size. 
$G \in SU(N_c)$, and the $J^a$ are the generators of the 
$SU(2)$ subgroup 
obtained by the upper-left-hand corner embedding
of the two-dimensional representation of $SU(2)$ into the 
$N_c$-dimensional representation of $SU(N_c)$ [\BERWEI ]. 
Substituting the instanton solution 
into the gauge kinetic term 
of the action \lag, the action takes the value
$S_{g}={8\pi^2\over g^2}$.

The bosonic zero-modes depend on the instanton configurations,
the size $\rho$, the location $x_0$ and the freedom of the embedding $G$.
{}From the gauge invariance, 
this embedding is determined by the $SU(N_c)$ rotation of
the scalar vacuum expectation values $\Omega$ in \expval.
Hence, fixing  $G=1$,
the integration measure for the bosonic degrees of freedom is 
given by [\BER,\COR]
$$
{2^{4{N_c}+2}\pi^{4{N_c}-2}\mu^{4N_c}\over ({N_c}-2)! ({N_c}-1)!g^{4N_c}}
\int d\rho \rho^{4N_c-5} \int d\Omega \int d^4x_0,
\eqn\bosmea
$$
where the group integration is normalized as
$\int d\Omega=1$.
Here we have introduced the regulator mass $\mu$.

Let us consider the fermionic part. 
Using supersymmetry and superconformal symmetry, 
the gauginos have the following 
zero-modes: 
$$
\eqalign{
{{\lambda^{SS}}_{\alpha a}}^b&=-{\sqrt{2}\rho^2 \over \pi}
 {x_\mu x_\nu \eta_{c\nu\lambda}\bar{\eta}_{d\mu\lambda} 
(\tau^c \xi_{SS})_\alpha
 \over x^2(x^2+\rho^2)^2}{ ({J^d})_a}^b, \cr
{{\lambda^{SC}}_{\alpha a}}^b&=
{\rho\over \pi} {x_\nu\bar{\eta}_{c\nu\lambda}
(\tau_\lambda^+ \xi_{SC})_\alpha 
\over (x^2+\rho^2)^2} {({J^c})_a}^b,
\cr
{\lambda_{\alpha a}}^b&={\rho i\over \sqrt{2} \pi}
{x_\mu(\tau_\mu^-)_{ac}\varepsilon_{c\alpha}\xi^b \over
(x^2+\rho^2)^{3/2}\sqrt{x^2}},\cr
{{\bar{\lambda}}_{\alpha a}}^{\ \ \ b}&={\rho i\over \sqrt{2} \pi}
{x_\mu (\tau_\mu^+)_{\alpha b} \bar{\xi}_a
\over (x^2+\rho^2)^{3/2}\sqrt{x^2}},\cr
}
\eqn\ferzermod
$$
where we have introduced the Grassmann odd numbers ${\xi_{SS}}^\alpha$,
${\xi_{SC}}^\alpha$, $\xi^a$ and $\bar{\xi}_a$ 
$(\alpha=1,2,\  a=3,\cdots, N_c)$  
to label the gaugino zero-modes.
$\varepsilon_{ab}$ denotes the
2 by 2 anti-symmetric tensor with $\varepsilon_{12}=1$, and
the $\tau_\mu^+$ and $\tau_\mu^-$ denote $(\sigma,-i)$ and $(\sigma,i)$,
respectively.
The matter zero-modes are given by the similar 
expressions as \ferzermod, and 
we use $\zeta$ instead of $\xi$ to label them.
The measure for the fermionic zero-modes is given by
$
\int d^{2{N_c}}\xi d^{2{N_c}}\zeta \mu^{-2{N_c}}
$
[\THO].

Now we turn on the vacuum expectation values of the 
scalar fields.
The instanton solution is not an exact solution anymore.
But if the instanton size $\rho$ is much smaller than the 
vacuum expectation values, the field equations are well approximated
by\footnote{\star}{See [\ADS] and [\AFF] for more detailed discussions}
$D_\mu G_{\mu\nu}=0$ and $D^2A=0$.
Thus the instanton solution 
remains as the solutions in this approximation.
In order to discuss the solutions for scalar fields, 
it is convenient to  divide the row and column of the 
the scalar field into the following blocks;
$$
A=
\left( \matrix{ 
A^{(1)} & A^{(2)} \cr
A^{(3)} & A^{(4)}
}\right),
\eqn\defconsca
$$
where $A^{(1)}$, $A^{(2)}$, $A^{(3)}$ and $A^{(4)}$ are 
$2\times2$, $2\times (N_{c}-2)$,$(N_{c}-2)\times 2$ and 
$(N_{c}-2)\times (N_{c}-2)$ matrices, respectively.
Then, 
the solutions for the scalar fields are given by
$$
\eqalign{
A^{(1)}(x)&={x^2\over x^2+\rho^2}\< A\>^{(1)}_{tl}
+{1\over 2} {\rm Tr}(\< A\>^{(1)})I_2, \cr
A^{(2,3)}(x)&=\sqrt{x^2\over x^2+\rho^2}\< A\>^{(2,3)}, \cr
A^{(4)}(x)&=\< A\>^{(4)},\cr
}
\eqn\scasol
$$
where $I_2$ is the $2\times 2$ identity matrix,
and $\< \cdot\>_{tl}$ denotes the traceless part.

The classical matter action under the non-vanishing 
scalar expectation values is given by substituting the kinetic term of 
the matter in \lag\ with the approximate classical solutions \scasol\
and the instanton solution. 
We thus obtain   
$$
\eqalign{
S_m&=8\pi^2\rho^2f,\cr
f&=
{\rm Tr}\left(\< A^\dagger\>_{tl}^{(1)}\< A\>_{tl}^{(1)}
+{1\over 2}\< A\> ^{(3)}\< A ^{\dagger}\> ^{(2)} +{1\over 2}
\< A\> ^{(2)}\< A^{\dagger}\>^{(3)}\right).\cr
}
\eqn\clavalact
$$

When the scalars have non-vanishing expectation values, the 
zero-modes other than the supersymmetric ones 
form Dirac mass terms
$2 (\bar{\zeta}_{SC},\zeta,\bar{\zeta})M
(\bar{\xi}_{SC},\xi,\bar{\xi})^t$
through the Yukawa-coupling
$\int d^4x 2\sqrt{2} g i$ ${\rm Tr}(\psi[A^\dagger ,\lambda])$.
By substituting the fermionic zero-modes \ferzermod\ and
the solutions of the scalar fields \scasol\ into the 
Yukawa-coupling, we obtain the following
mass matrix $M$:
$$
\eqalign{
M&={gi\over 2}\left(\matrix{
\sqrt{2}\varepsilon \< A^\dagger\>_{tl}^{(1)} & 
(\< A^\dagger\>^{(3)})^t & 
\varepsilon \< A^\dagger\> ^{(2)} \cr
\< A^\dagger\> ^{(3)} & 0 & 
-{{\rm Tr} \< A^\dagger\> ^{(1)} \over \sqrt{2}} I_{N_{c}-2}
+ \sqrt{2}\< A^\dagger\> ^{(4)} \cr
(\varepsilon \< A^\dagger\> ^{(2)})^t & 
-{{\rm Tr}\< A^\dagger\> ^{(1)}\over \sqrt{2}} I_{N_{c}-2}
+\sqrt{2}(\< A^\dagger\> ^{(4)})^t 
&0\cr
}\right),\cr
}
\eqn\masmat
$$
where $I_{N_{c}-2}$ denotes the $(N_c-2)\times (N_c-2)$ identity matrix. 

The integration over the non-supersymmetric  
zero-modes gives the contribution of the determinant of the
mass matrix: ${\rm det}(2M)$.
Since there remain the two supersymmetric zero-modes of each 
matter fermion and gaugino field, 
we must insert two matter fermion and two gaugino fields
into a correlation function.
The appropriate choice of the fields is given by the classically 
massless fermions $\psi^\dagger_0(x)={\rm Tr}(\< A\> \psi^\dagger (x))$,
and similarly for $\lambda^\dagger_0(x)$.
Since the Yukawa couplings with the non-vanishing scalar expectation
values mix $\lambda$ and $\psi$, 
the surviving supersymmetric zero  modes $\lambda^{SS}$ and
$\psi^{SS}$ smear to $\psi^\dagger$ and $\lambda^\dagger$,
respectively.
To the first order in $\rho A$, the smearing to $\psi^\dagger$
is estimated by the classical field equation of $\psi^\dagger$:
$\sigma^\mu D_\mu ^* \psi^\dagger-\sqrt{2}g[A^\dagger,\lambda^{SS}]=0.$
A similar equation can be derived for $\lambda^\dagger_0(x)$,
and we obtain, for $x-x_0\gg \rho$,
$$
\psi^\dagger_0(x_1)\psi^\dagger_0(x_2)
\lambda^\dagger_0(x_3)\lambda^\dagger_0(x_4) 
\sim
(\pi g\rho^2 f)^4 
\left({1\over 2} \xi^2_{SS}\right)\left({1\over 2}\zeta^2_{SS}\right)
S_F^4,
\eqn\foufer
$$
where we have introduced the notation 
$
S_F^4=S^{\alpha\beta}_F(x_1-x_0)S_{F
\alpha\beta}(x_2-x_0)S^{\alpha'\beta'}_F(x_3-x_0)
S_{F\alpha'\beta'}(x_4-x_0)
$
with $S_F$ and $x_0$ denoting the fermion propagator and 
the location of the instanton, respectively. 
The insertion of \foufer\ will eliminate the remaining supersymmetric
zero-modes in the fermionic integration measure.

After the integration over the fermionic zero-modes
and the size of the instanton $\rho$, we obtain
the four point function
$$
\< 
\psi^\dagger_0(x_1)\psi^\dagger_0(x_2)\lambda^\dagger_0(x_3)
\lambda^\dagger_0 (x_4) \> 
={\Lambda_{N_c}^{2{N_c}} \Gamma(2{N_c}+2) \over 
2^7 \pi^2 g^{2{N_c}-2}({N_c}-2)! ({N_c}-1)!}
\int d\Omega {{\rm det}M \over (gf)^{2{N_c}-2}}
\int d^4 x_0 S_F^4,
\eqn\parint
$$
where we have introduced the dynamical scale 
$\Lambda_{N_c}^{2N_c}=\mu^{2N_c}\exp({-8\pi^2/g^2})$.

Now we discuss the integration over the group manifold $\int
d\Omega$.
The integration domain can be enlarged from $SU(N_c)$
to $U(N_c)$ trivially.
Also one can easily see, 
{}from the explicit definition of $M$ and $f$ ,
that the integrand has the symmetries of $U(2)\times U(N_c-2)$, 
where the $U(2)$ is the unitary adjoint rotation of the
upper-left-hand corner, and the $U(N_c-2)$ is that of the bottom-right-hand
corner.
Hence the integration to be done is over the Grassmannian manifold
$U(N_c)/(U(2)\times U(N_c-2))$.
This integration looks very complicated. 
We will show that  the holomorphy argument [\SEIONE,\AMAKON,\SHIVAI]
simplifies the integration.

We will begin with the simplest case $N_c=3$.
The parameterization of the Grassmannian manifold and the measure are
given by [\COR]
$$
\eqalign{
\Omega&=\left(\matrix{(1-yy^{\dag})^{1/2} & -y \cr y^{\dag} & 
(1-y^{\dag}y)^{1/2}}\right), \cr
y&=\left(\matrix{y_1\cr y_2}\right),\cr
\int d\Omega &= 2 \int_{0<r_1,r_2}^{r_1+r_2<1} dr_1dr_2,
}
\eqn\intmea
$$
where we have used the fact that the integrand depends on the
$y_1$ and $y_2$ only through $r_1=\mid y_1 \mid^2$, $r_2=\mid y_2 \mid^2$,
because there remain $U(1)^2$ symmetry for the flat direction \expval.

Parameterizing the scalar vacuum expectation values as $a_1=v-w$,
$a_2=-v-w$ and $a_3=2w$, we get an explicit form for the integrand
${\rm det}M/(gf)^4$. 
{}From the holomorphy argument, the result of the
integration must be independent of the conjugate fields $v^*$ and 
$w^*$.
Physically, we must regard $v^*$ and $w^*$ as the complex 
conjugates of $v$ and $w$, but mathematically, by the 
analytic continuation, we can regard $v^*$ and $w^*$
as the variables independent of $v$ and $w$.
Thus we have the freedom to set the values of $v^*$ and $w^*$ to 
make the integrands as simple as possible.

One of choices is to set $w^*=0$.
In this case ${\rm det}M$ and $f$ become
$$
\eqalign{
{\rm det}M=
&-{g^4{v^*}^4 \over 64} (
16 r_1^2  - 8 r_1^3  + r_1^4  - 16 r_1
r_2  + 8 r_1^2 r_2  - 4 r_1^3 r_2  
+ 16 r_2^2  + 8 r_1 r_2^2 \cr
& + 6 r_1^2 r_2^2  
- 8 r_2^3  - 4 r_1 r_2^3  + r_2^4 ), \cr
f=&2v^*
(4v-2vr_1-(v+3w)r_1^2-2vr_2+2vr_1r_2-(v-3w)r_2^2).
\cr
}
\eqn\mfxdef
$$
The indefinite integration over $r_1$ and $r_2$ can be performed 
explicitly with the elementary functions.
Thus we obtain
$$
\int  d\Omega {{\rm det}M \over (gf)^4}=
-{v^2+3w^2\over 16v^2(9w^2-v^2)^2}.
\eqn\intres
$$
Another choice is to set $v^*=0$. In this case, the vacuum
expectation values of the conjugate scalar field has an enhanced
symmetry of $SU(2)$, and the integrand becomes simpler.
By setting naively $v^*=0$, we obtain
$$
{{\rm det}M \over (gf)^4}=
-{1\over 4(6w+(v-3w)r_1-(v+3w)r_2)^4}.
\eqn\naiset
$$
Since the denominator of \naiset\ consists  of a linear function 
of $r_1$ and $r_2$,
the integration seems to be easily done. 
But one finds that this naive integration does not give the correct
answer, since the procedures of setting $v^*=0$ and the integration do not
commute at the origin $r_1=r_2=0$.
By blowing up near the origin, one finds 
that there is a delta functional contribution from the origin:
$$
2 \int_{0<r_1,r_2}^{r_1+r_2<+0}dr_1dr_2 {{\rm det}M \over (gf)^4}
=-{1\over 2^43^3v^2w^2}.
\eqn\delcon
$$
Adding the integration of \naiset\ to \delcon, we again obtain the 
result \intres.

Since the denominator of the result \intres\ is proportional to 
the discriminant $\triangle_3=\prod_{i<j}(a_i-a_j)^2$,
the result diverges when two of the $a_i$ have the same value.
This condition of the divergence is characterized by 
$f=0$. 
In fact, if we assume $f=0$ and 
take $\< A^\dagger\> = \< A\>^\dagger$,
we obtain 
two of the $a_i$'s  must take the same value.

Now we evaluate the order and the numerical factor of 
the divergence for the general case of $SU(N_c)$. 
Let us introduce an infinitesimal parameter $\epsilon$, and 
parameterize the vacuum expectation values of the scalar field as follows:
$$
\eqalign{
\< A\>&=\Omega A_0 \Omega^\dagger,\cr
A_0&=\left( \matrix{{A_0}^{(1)}&0\cr 0&{A_0}^{(4)}}\right),\cr
{A_0}^{(1)}&=\left( \matrix{a+\epsilon&0\cr 0&a-\epsilon} \right), \cr
\< A^\dagger\> &=\Omega \bar{A_0} \Omega^\dagger.
}
\eqn\vacexpsca
$$
where ${A_0}^{(4)}$ and $\bar{A_0}$ are diagonal matrices.
But here the conjugate scalar expectation values $\bar{A_0}$   
are taken to be independent of the $A_0$, 
and we assume
the diagonal elements of $\bar{A_0}$ take different 
values from each other.
One notices that the vacuum expectation values \vacexpsca\ 
are the flat directions.

Since $f=\epsilon (\bar{a}_1-\bar{a}_2)$ for $\Omega=1$, the important 
contribution of the integration will come from the $\Omega$
which gives $f\sim O(\epsilon)$.
Hence we blow up $\Omega$ as
$$
\Omega=\left( \matrix{U_1&\sqrt{\epsilon}U_2\cr 
\sqrt{\epsilon}U_3&U_4}\right).
\eqn\forome
$$
Substituting \forome\ into \vacexpsca\ and 
using the unitarity of the $\Omega$, we obtain
$$
\eqalign{
f=&\epsilon \Bigl( \bar{a}_1-\bar{a}_2+{\rm Tr}
\left({A_0}^{(4)}\bar{A_0}^{(4)}W^\dagger W\right)
-a{\rm Tr}\left(W^\dagger W\bar{A_0}^{(4)}\right) \cr
&-{1\over 2}{\rm Tr}\left(\bar{A_0}^{(1)}\right)
{\rm Tr}\left({A_0}^{(4)}W^\dagger W\right) 
+{a\over 2}{\rm Tr}\left(W^\dagger W\right) 
{\rm Tr}\left(\bar{A_0}^{(1)}\right) \Bigr) 
+O(\epsilon^2),
}
\eqn\appfep
$$
where $W=U_1^\dagger U_2$.

Since ${\rm det}M$ depends only on $\< A^\dagger\>$, the 
lowest order of ${\rm det}M$ is $O(\epsilon^0)$.
The 0-th order of $\Omega$ in $\epsilon$ 
is an element of the invariant group $U(2)\times U(N_c-2)$.
Hence it is enough to evaluate ${\rm det}M$ under $\Omega=1$,
and we obtain
$$
{\rm det}M=-{1\over 4} \left({g^2\over 2} \right)^{N_c-1}
(\bar{a}_1-\bar{a}_2)^2
\prod_{i=3}^{N_c}\left( \bar{a}_i-{\bar{a}_1+\bar{a}_2\over 2}\right)^2.
\eqn\detzer
$$

The integral measure is explicitly defined by
$$
\eqalign{
 d\Omega &={1\over V_{N_c}}
\prod_{i=1,j=1}^{N_c}d^2\Omega_{ij}
\delta(\Omega\Omega^\dagger-1),\cr
\delta(\Omega\Omega^\dagger-1)&=
\prod_{i=1}^{N_c}\delta((\Omega\Omega^\dagger)_{ii}-1)
\prod_{j<k}^{N_c}\delta^2((\Omega\Omega^\dagger)_{jk}),\cr
}
\eqn\intomedef
$$
where $V_{N_c}$ is a normalization constant.
Using the result in the literature [\BER],
this volume factor is obtained as 
$V_{N_c}=\pi^{N_c(N_c+1)\over 2} \prod_{k=2}^{N_c} {1\over (k-1)!}$.
Substituting \forome\ into \intomedef, we obtain, in the lowest order of 
$\epsilon$,
$$
\int  d\Omega {{\rm det}M\over (gf)^{2{N_c}-2}}
= \epsilon^{2(N_c-2)}{V_2V_{N_c-2}\over V_{N_c}}\int dW
{{\rm det}M\over (gf)^{2{N_c}-2}},
\eqn\expgroint
$$
where we have integrated over the invariant group 
$U_1\times U_4 \in U(2)\times U(N_c-2)$,
and have changed the integration variable $U_2$ by the 
transformation $W=U_1^\dagger U_2$.

To evaluate the integration, let us parameterize 
$$
W=\left(\matrix{ y_3 & \cdots &y_{N_c} \cr z_3 & \cdots & z_{N_c}}
\right).
\eqn\parw
$$
Then we get
$$
\eqalign{
f&=\epsilon\left( \bar{a}_1-\bar{a}_2+\sum_{i=3}^{N_c} 
\left( \bar{a}_i-{\bar{a}_1
+\bar{a}_2\over 2}\right) (a_i-a)(\mid y_i \mid^2+
\mid z_i \mid^2)\right),
\cr
dW&=\prod_{i=3}^{N_c} \pi^2 d\mid y_i\mid^2 d\mid z_i\mid^2.
\cr
}
\eqn\expexpfw
$$
Thus we finally obtain
$$
\int d\Omega {{\rm det}M \over (gf)^{2N_c-2}}\sim 
-{(N_c-1)!(N_c-2)! \over 2^{N_c+1}(2N_c-3)!}
{1\over \epsilon ^2 \prod_{i=3}^{N_c}(a_i-a)^2}.
\eqn\finres
$$
for $a_1=a+\epsilon$ and $a_2=a-\epsilon$.
Here the dependences on $\bar{a}_i$ have certainly disappeared. 

{}From the gauge invariance, the full expression must be a 
symmetric function of $a_i$.
If we also assume the full expression is a rational function
of $a_i$, we obtain, by a dimensional consideration, a unique result
$$
\int d\Omega {{\rm det}M\over (gf)^{2{N_c}-2}}
=-{({N_c}-1)!({N_c}-2)! \over 2^{N_c} (2{N_c}-3)!} 
{\triangle'_{N_c}(a) \over \triangle_{N_c}(a)},
\eqn\finresgen
$$
where
$$
\eqalign{
\triangle_{N_c}(a)=&\prod_{k<l}^{N_c}(a_k-a_l)^2, \cr
\triangle_{N_c}'(a)=&\sum_{i=1}^{N_c} 
\prod_{k<l,k\neq i,l\neq i}^{N_c}(a_k-a_l)^2, 
\cr 
\Bigl( \triangle_{N_c}'(a) &=2\ \ 
{\rm for}\  N_c=2 \Bigr) .
\cr
}
\eqn\tridef
$$
Note that  the present formula  \finresgen\ for  $N_c=3$ 
agrees with \intres.

Substituting \finresgen\ into \parint, we finally obtain
the four point function
$$
\langle \psi^{\dag}_0(x_1) 
\psi^{\dag}_0(x_2) \lambda^{\dag}_0(x_3) 
\lambda^{\dag}_0(x_4) \rangle 
=-{\Lambda_{N_c}^{2{N_c}} \Gamma(2{N_c}+2) \over 
2^{{N_c}+7}\pi^2  g^{2{N_c}-2} \Gamma(2{N_c}-2)} 
{\triangle'_{N_c}(a) \over \triangle_{N_c}(a)}
\int d^4x_0 S_F^4
\eqn\finresfou
$$

Now we will derive the 1-instanton correction of the prepotential
{}from the result \finresfou.
In the $N=1$ language, 
the $N=2$ $SU(N_c)$ SYM theory 
has the following low energy effective Lagrangian density:
$$
\eqalign{
{\cal L}&={1\over 4\pi} {\rm Im}\left[ 
\int d^4\theta \left({\partial {\cal F}(\phi)\over \partial \phi_i} 
\bar{\phi}_i\right) + 
\int d^2\theta {1\over 2} \left({\partial^2{\cal F}(\phi)
\over \partial \phi_i
\partial \phi_j} W^\alpha_{\ i}W_{\alpha j} \right) \right]
\cr
&={2\over g^2}\int d^4\theta \left( {\rm Tr}\phi_i\bar{\phi}_i\right)
+{1\over 8\pi}{\rm Im}\int d^2\theta \left({\partial^2 {\cal F}_1
\over \partial \phi_i \partial \phi_j} W^\alpha_{\ i}W_{\alpha j} \right)
+ \cdots,
}
\eqn\lagpre
$$
where ${\cal F}_1$ denotes the 1-instanton correction to the 
prepotential.

The 1-instanton correction to the four point function is given by
$$
\langle \psi^{\dagger}_0(x_1)
\psi^{\dagger}_0(x_2) 
\lambda^{\dagger}_0(x_3) 
\lambda^{\dagger}_0(x_4) \rangle 
=
{g^8\over 2^7\pi i}
\< a_j\>\< a_k\>\< a_l\>\< a_m\> {\partial^4{\cal F}_1 \over
\partial a_j\partial a_k\partial a_l\partial a_m}
\int d^4x_0 S_F^4.
\eqn\finresexa
$$
Comparing with the 1-instatanton calculation \finresfou\ and 
taking into account the rescaling of the fields by $g$
\footnote{\star}{The normalizations of the kinetic terms are different
by $g$ between \lag\ and \lagpre }, we obtain
$$
{\cal F}_1=-{i \Lambda_{N_c}^{2{N_c}} \over 2^{N_c} \pi} 
{\triangle'_{N_c}(a) \over \triangle_{N_c}(a)}.
\eqn\finresone
$$

One can check this result \finresone\ 
by using consistency under the 
matching condition of the gauge coupling. 
Parameterize  the scalar vacuum expectation values as
$$
\eqalign{
a_i&=a'_i-b\ \ (i=1,\cdots,N_c-1), \cr
a_{N_c}&=(N_c-1)b, \cr
}
\eqn\parsca
$$
and consider $b$ is very large compared to $a'_i$.  
Then, below the scale $b$, the gauge group of the system is 
effectively $SU(N_c-1)$ with the dynamical scale given by the 
following matching condition [\FINPOU]:
$$
m^2\Lambda_{N_c-1}^{2(N_c-1)}=\Lambda_{N_c}^{2N_c}.
\eqn\matcon
$$
Here the matching scale $m$ is the mass of the gauge bosons
of the gauge symmetry broken by the $b$,
and its value is $m^2=2(N_cb)^2$.
On the other hand, substituting \parsca\ into \finresone, we obtain,
in the large $b$ limit,
$$
{\cal F}_1= 
-{i \Lambda_{N_c}^{2{N_c}} \over 2^{N_c} \pi (N_cb)^2} 
{\triangle'_{N_c-1}(a') \over \triangle_{N_c-1}(a')}
+o(b^{-2}).
\eqn\onelim
$$
Using the matching condition \matcon, this is nothing but the 
1-instanton correction \finresone\ with the gauge group $SU(N_c-1)$.

A further check comes from the exact solutions discussed recently.
The discussions are based on the hyperelliptic curve [\KLELERYAN,\ARGFAR]
$$
y^2=\prod_{i=1}^{N_c}(x-e_i)^2-(\Lambda_{N_c}^{\rm ex})^{2N_c},
\eqn\hypell
$$
where $a_i\sim e_i$ classically.
Taking the $b$ very large and rescaling the $x$ and $y$
appropriately, one obtains the following matching condition[\ARGFAR]:
$$
({\Lambda_{N_c}^{\rm ex}})^{2N_c}
=(N_cb)^2({\Lambda_{N_c}^{\rm ex}})^{2(N_c-1)}
\eqn\lamexamat
$$
Comparing with the physical matching condition \matcon, 
the both scales should have the following relation: 
$$
\Lambda_{N_c}^{\rm ex}=2^{d-N_c \over 2N_c}\Lambda_{N_c},
\eqn\lamrel
$$
where $d$ is a parameter to be fixed.
Substituting \lamrel\ into \finresone\ and \lagpre,
we obtain
$$
{\cal F}
=i\left({4\pi \over g_0^2}-{N_c\over 2\pi}
{\rm ln}({\Lambda_{N_c}^{\rm ex}})^2+\cdots \right){\rm Tr}(\phi^2)
-{i ({\Lambda_{N_c}^{\rm ex}})^{2{N_c}} \over 2^{d} \pi} 
{\triangle'_{N_c}(a) \over \triangle_{N_c}(a)}+\cdots,
\eqn\coroneexa
$$
where we have explicitly written down the dynamical scale dependence
of the one-loop correction to compare the prepotential \coroneexa\ 
with the exact solutions unambiguously\footnote{*}{Actually, the 
normalization of the prepotential depends on literatures.}

The explicit expressions of the one-instanton corrections of 
the exact prepotentials for the $SU(2)$ and
$SU(3)$ cases are known [\SEIWITONE,\KLELER]:
$$
\eqalign{
{\cal F}^{\rm ex}_{SU(2)}&=i\left( -{1\over 4\pi}{\rm ln}
({\Lambda_2^{\rm ex}})^2+\cdots
\right){\rm Tr}(\phi^2) -{i({\Lambda_2^{\rm ex}})^4 \over 2^4 \pi}
{\triangle'_{2}(a) \over \triangle_{2}(a)}+\cdots,\cr
{\cal F}^{\rm ex}_{SU(3)}&=i\left(-{3\over 4\pi} 
{\rm ln}({\Lambda_{3}^{\rm ex}})^2+\cdots\right){\rm Tr}(\phi ^2)
-{i({\Lambda_{3}^{\rm ex}})^6 \over 2^3 \pi} 
{\triangle'_{3}(a) \over \triangle_{3}(a)}+\cdots.
}
\eqn\KLTexa
$$
Normalizing with the one-loop coefficients, our result \coroneexa\ 
with $d=2$ agrees with the exact solutions \KLTexa\ for the both cases.
Further this $d=2$ agrees with the discussions by Finnell and Pouliot 
[\FINPOU], 
who derived the similar relation between the scales for the $N_c=2$ case 
by discussing the physical matching condition of the gauge couplings 
between the original and the effective theories.

We have explicitly performed the 1-instanton calculation and 
derived the 1-instanton correction to the prepotential
{}from the microscopic point of view.
Although there appeared a difficulty in the  integration over the 
Grassmannian 
manifold $U(N_c)/(U(2)\times U(N_c-2))$, we have done
this integration by using the holomorphy argument effectively.
The present method would be applicable to other gauge groups
and the $N=2$ massive QCD.
Concerning the higher instanton contributions,
it seems a quite interesting problem 
to calculate them from the microscopic 
approach and compare them with the exact solutions.
We note that the holomorphy argument holds for general $N=1$ supersymmetric
theories. 
Therefore our method will be effective for other $N=1$ models.

\ack

We would like to thank S.-K.~Yang for stimulating discussions
and encouragement, and N.S. would also like to thank the elementary particle
group of Tohoku University and especially S.~Watamura
for their hospitality.
The work of K.I. is supported in part by University of Tsukuba
Research Projects and the Grant-in-Aid for Scientific Research from
the Ministry of Education (No.~07210210).
The work of N.S. is supported by the JSPS fellowship  and the
Grant-in-Aid for Scientific Research from
the Ministry of Education (No.~06-3758).

\refout

\bye